\documentclass[11pt]{iopart}

\newcommand{\lagrangian}{\mathcal{L}}
\newcommand{\M}{\mathcal{M}}
\newcommand{\orbit}{\mathcal{O}}
\newcommand{\Lie}{\mathcal{L}}
\begin{document}
\begin{flushright}
 OCU-PHYS 316, AP-GR 68
\end{flushright}

\title{Exactly solvable strings in Minkowski spacetime}

\author{Hiroshi Kozaki$^1$, Tatsuhiko Koike$^2$ and Hideki Ishihara$^3$}

\address{$^1$ Department of General Education, Ishikawa National College
of Technology, Tsubata, Ishikawa 929-0392, Japan}

\address{$^2$ Department of Physics, Keio University, Yokohama 223-8522,
Japan}

\address{$^3$ Department of Mathematics and Physics,
Graduate school of Science, Osaka City University,
Osaka 558-8585, Japan}

\eads{\mailto{kozaki@ishikawa-nct.ac.jp},
 \mailto{koike@phys.keio.ac.jp},
 \mailto{ishihara@sci.osaka-cu.ac.jp}}

\begin{abstract}
 We study the integrability of the equations of motion for the
 Nambu-Goto strings with a cohomogeneity-one symmetry in Minkowski
 spacetime.
 A cohomogeneity-one string has a world surface which is tangent to a
 Killing vector field. By virtue of the Killing vector,
 the equations of motion reduce to the geodesic equation in the
 orbit space.
 Cohomogeneity-one strings are classified into seven 
 classes (Types I to VII). We investigate the integrability of the
 geodesic equations for all the classes and find that the geodesic
 equations are integrable. 
 For Types I to VI, the integrability comes from the existence of
 Killing vectors on the orbit space 
 which are the projections of Killing vectors
 on Minkowski spacetime.
 For Type VII, the integrability is related to a projected Killing
 vector and a nontrivial Killing tensor on the orbit space.
 We also find that the geodesic equations of all types are exactly
 solvable, and show the solutions.
\end{abstract}


\section{Introduction}
Cosmic strings are topological defects which are produced when the
$U(1)$ symmetry breaks down in the unified theories. Such a symmetry
breaking is supposed to have occurred at the early stage of the
universe.
If the existence of the cosmic strings is confirmed,
it is a strong evidence of vacuum phase transition in the universe.
Besides the interests from the unified theories, 
cosmic strings have been studied in the context of cosmology since
they were proposed as possible seeds of the structure formation in the
universe \cite{Zeldovich:1980gh, Vilenkin:1981iu}. 
However, this scenario was rejected due to the confliction with the
precise observational data of cosmic microwave backgrounds
\cite{Albrecht:1997nt, Albrecht:1997mz}.

Recently, cosmic strings gather much attention in the context of the
superstring theories since fundamental strings and other string-like
solitons such as D-strings could exist in the universe as cosmic strings
\cite{Copeland:2003bj}. 
In the brane inflation models, these cosmic superstrings are produced at
the end of inflation \cite{Sarangi:2002yt,Dvali:2003zj} and stretched by
the expansion of the universe.
Detection of the cosmic superstrings will give a strong evidence of the
superstring theories.

One of the main difference between the gauge-theoretic cosmic strings
and the cosmic superstrings is in the reconnection probabilities.
For the former strings, the reconnection probability is almost one.
Therefore, when a long gauge-theoretic string intersects with itself,
it breaks up to a closed string and an open string which is
shorten by the reconnection.
Closed strings cannot exist stably since they decay by radiating
gravitational waves. 
Then, most of the gauge-theoretic strings could not survive in the
universe.
For the cosmic superstrings, it is clarified that the reconnection
probability may be much suppressed \cite{Jackson:2004zg}.
If it is true, the cosmic superstrings can survive in the universe and
stay in isolation.
Since they rarely interact with each other, 
they must have gone through some relaxation process such as
gravitational radiations, and then, they must be in stationary motions.

The stationary string has a world surface which is tangent to a timelike
Killing vector field. 
Existence of the tangent Killing vector reduces the equations of
motion to ordinary differential equations, 
which are much more tractable than partial differential equations. 
Stationary strings in various spacetimes have been studied so
far \cite{Burden:1982zb, Burden:1984xk, Frolov:1988zn, 
deVega:1993rm, Larsen:1994ah, Larsen:1995bp, deVega:1996mv,
Frolov:1996xw, Kubiznak:2007ca, Ogawa:2008qn, Koike:2008fs, Burden:2008zz}, 
and many non-trivial solutions were found even in Minkowski
spacetime \cite{Burden:1982zb, Burden:1984xk, Frolov:1988zn,
Frolov:1996xw, Ogawa:2008qn, Burden:2008zz}.

The notion of stationary string is generalized to that of
cohomogeneity-one string.
A cohomogeneity-one string is defined as a string whose world surface
is tangent to a Killing vector field on the spacetime 
which is not restricted to being
timelike. If the Killing vector is timelike, the cohomogeneity-one
string is stationary.
The cohomogeneity-one string is characterized by the tangent Killing
vector.  When the spacetime admits multiple
independent Killing vectors, there are infinitely many Killing
vectors in the form of linear combinations. 
Correspondingly,
infinitely many cohomogeneity-one strings are possible.
However, we do not have to distinguish all of them.
For example, 
in Minkowski spacetime,
we can identify two stationary rotating strings which have
equal angular velocity and different rotational axes, e.g. $x$-axis and
$y$-axis.
This identification is generalized as follows:
two strings are equivalent if their world surfaces, say $\Sigma_1$
and $\Sigma_2$, are mapped by an isometry $\varphi$, 
\begin{equation}
  \Sigma_2=\varphi(\Sigma_1) . 
\end{equation}
In the case of the cohomogeneity-one strings, we can identify the
strings if the isometry $\varphi$ sends 
the tangent Killing vector $\xi_1$ 
which defines the cohomogeneity-one
property of $\Sigma_1$ 
to the Killing vector $\xi_2$ which defines that of $\Sigma_2$: 
\begin{equation}
  \xi_2=\varphi_{\ast}\xi_1. 
\end{equation}

In Minkowski spacetime, the Killing vectors are classified into
seven families (Types I to VII) under identification by 
isometries.
Therefore, the cohomogeneity-one strings fall into seven
families \cite{Ishihara:2005nu}. 
The Type I family includes stationary rotating strings. 
The Nambu-Goto equations of motion for this class are exactly solved and
various configurations are found
\cite{Burden:1982zb, Burden:1984xk,Burden:2008zz, 
Frolov:1988zn,Frolov:1996xw,Ogawa:2008qn}.
By using the exact solutions, the energy momentum tensors
are calculated and the properties of the stationary rotating strings are
clarified \cite{Ogawa:2008qn}.
The gravitational perturbations are also studied in detail
and the wave form of the gravitational waves are obtained \cite{Ogawa:2008yx}.
The classification of the cohomogeneity-one strings corresponding to 
\cite{Ishihara:2005nu} is done also in the anti-de Sitter spacetime in five
dimensions~\cite{Koike:2008fs}. 

The studies of Type I family show that exact solutions are
useful for investigating the gravitational phenomena such as 
gravitational lensing and gravitational waves,  
which are indispensable to verifying the existence of cosmic strings.
Furthermore, exact solutions provide us with a deeper insight to the
string dynamics. 
In this paper, we clarify the integrability of the remaining families
(Types II to VII) to complete exact solutions of cohomogeneity-one
strings in Minkowski spacetime. 
We assume that the motions of the strings are governed by the Nambu-Goto
action.

We emphasize that the integrability above means exact solvability (up to
quadrature) of the embedding of the string's world surface into the spacetime
and should be clearly distinguished from the well-known ``integrability'' 
of classical strings which means existence of infinite number of conserved
quantities.
To be concrete, when we work in the conformal gauge, the Nambu-Goto
equations in Minkowski spacetime reduce to the wave equations in the
two-dimensional flat spacetime supplemented with constraint equations.
Though general solutions of the former can be easily obtained,
the latter are not solvable in general, because they are non-linear partial 
differential equations. In the case of cohomogeneity-one strings,
the equations of motion reduce to geodesic equations in a curved space, 
where the metric of the reduced space depends on the symmetry of the world
surface \cite{Koike:2008fs,Ishihara:2005nu}
Even in this case, however, it is still non-trivial whether
the geodesic equations on the metric are integrable or not.

In the next section, we review the equations of motion for the
cohomogeneity-one strings and argue the integrability.
In Sec. \ref{sec:solutions}, we solve the equations of motion and give 
the solutions in the closed forms.
We conclude the work in the final section.

\section{Integrability of cohomogeneity-one strings}\label{sec:c1string}

A trajectory of the string is a two-dimensional surface, say $\Sigma$,
embedded in the spacetime $(\M, g$).
We denote the embedding as
\begin{equation}
 \zeta^a \mapsto x^{\mu} = x^{\mu}(\zeta^a),
\end{equation}
where $\zeta^a(\zeta^0 = \tau, \zeta^1 = \sigma)$ are the coordinate on
$\Sigma$ and $x^{\mu}~(\mu = 0, 1, 2, 3)$ are the coordinate in $\M$.
The Nambu-Goto action is written as
\begin{equation}
 S = \int_{\Sigma} \sqrt{- \gamma} d^2 \zeta,
\end{equation}
where $\gamma$ is the determinant of the metric $\gamma_{ab}$ 
induced on $\Sigma$ 
which is given by
\begin{equation}
 \gamma_{ab} = g_{\mu \nu} \frac{\partial x^{\mu}}{\partial \zeta^{a}}
                           \frac{\partial x^{\nu}}{\partial \zeta^{b}}.
\end{equation}

Let us consider the case that the spacetime $\M$ admits a Killing vector
field $\xi$. We denote the one-parameter isometry group generated by
$\xi$ as $H$. 
The group action of $H$ on $\M$ generates orbits of $H$, or the
integral curves of $\xi$.
If the string world surface $\Sigma$ is foliated by the orbits of $H$,
the string is called cohomogeneity-one associated with the Killing
vector $\xi$.
It is obvious that $\Sigma$ is tangent to $\xi$.

When we identify the points in $\M$ which are connected by the action
of $H$, we have the orbit space $\orbit := \M/H$.
Under this identification, the cohomogeneity-one world surface $\Sigma$
becomes a curve in $\orbit$.
This curve is shown to be a spacelike geodesic in $\orbit$ with the
norm-weighted metric 
\begin{equation}
\tilde{h}_{\mu \nu} = - f h_{\mu \nu}, 
\end{equation}
where $f$ is the squared norm of $\xi$ and $h_{\mu \nu}$ is a
naturally induced 
metric on $\orbit$:
\begin{equation}
 h_{\mu \nu} = g_{\mu \nu} - \xi_{\mu} \xi_{\nu} / f. 
 \label{eq:induced_metric}
\end{equation}
Therefore, the equations of motion for the cohomogeneity-one string
are reduced to the geodesic equations on $(\orbit, \tilde{h})$.

Integrability of the geodesic equations is related to the existence of
Killing vectors and Killing tensors.
Let $K^{\mu}$ be a Killing vector field on $(\orbit, \tilde{h})$,
 which satisfies the Killing equations 
\begin{equation}
 \nabla_{(\mu} K_{\nu )} = 0,
\end{equation}
where $\nabla_{\mu}$ denotes the covariant derivative 
with respect to $\tilde{h}$. 
For a tangent vector $u^{\mu}$ of a geodesic,
$K_{\mu} u^{\mu}$ is conserved along the geodesic.
Let $K_{\mu\nu}$ be a Killing tensor field on $(\orbit, \tilde{h})$,
which is symmetric and satisfies the Killing equations 
\begin{equation}
 \nabla_{(\mu} K_{\nu \lambda )} = 0,
\end{equation}
$K_{\mu\nu} u^{\mu} u^{\nu}$ is also conserved along the geodesic.
If the geodesic has enough number of such conserved quantities 
which commute with each other,
the geodesic equations are integrable.
In the case of geodesics in $(\orbit, \tilde{h})$,
where $\dim \orbit = 3$, two conserved quantities are required 
for the integrability.
Therefore, if $(\orbit, \tilde{h})$ admits two or more Killing vectors
and Killing tensors, geodesic equations are integrable

In the orbit space $(\orbit, \tilde{h})$, we can find such
Killing vectors without solving the Killing equations.
Let us consider a Killing vector $X$ in $(\M, g)$ which commutes with
$\xi$. We can easily find that the projection of $X$, 
say $\pi_{\ast} X$, where $\pi:\M \rightarrow \orbit$ is the projection,
is a Killing vector in $(\orbit, \tilde{h})$:

\begin{eqnarray}
 \Lie_{\pi_{\ast}X} \tilde{h}_{\mu \nu}
 &= \Lie_{X} (- f g_{\mu \nu} + \xi_{\mu}  \xi_{\nu})
  = \Lie_{X} \left\{%
               (%
                 - g_{\rho \sigma} g_{\mu \nu}
                 + g_{\mu \rho} g_{\nu \sigma}
               )
               \xi^{\rho}\xi^{\sigma}
             \right\} \nonumber \\
 &= (%
       - g_{\rho \sigma} g_{\mu \nu}
       + g_{\mu \rho} g_{\nu \sigma}
    )
    \left\{
      (\Lie_{X} \xi^{\rho}) \xi^{\sigma}
    + \xi^{\rho} \Lie_{X} \xi^{\sigma}
    \right\} \nonumber \\
 &= (%
       - g_{\rho \sigma} g_{\mu \nu}
       + g_{\mu \rho} g_{\nu \sigma}
    )
    \left\{
      [X, \xi]^\rho \xi^{\sigma}
    + \xi^{\rho} [X, \xi]^{\sigma}
    \right\} = 0.
\end{eqnarray}
Killing vectors which commute with $\xi$ constitute a Lie subalgebra,
called centralizer of $\xi$ which we denote ${\cal C}(\xi)$.
Let $X, Y \in {\cal C}(\xi)$ commute with each other.
We can show that the projections of them on $\orbit$ also commute;
\begin{equation}
 [\pi_{\ast}X, \pi_{\ast}Y] = \pi_{\ast}[X, Y] = 0.
\end{equation}
Therefore, if there are two or more linearly independent and commuting
Killing vectors in ${\cal C}(\xi)$ except for $\xi$ itself,
$(\orbit, \tilde{h})$ inherits the same number of commuting Killing
vectors, and then the geodesic equations in $(\orbit, \tilde{h})$ are
integrable.

In Minkowski spacetime, all of the cohomogeneity-one strings are
classified into seven families (Types I to VII).
For each type, we list the Killing vector $\xi$, 
basis of ${\cal C}(\xi)$ and the number of commuting basis of ${\cal C}(\xi)$ except
for $\xi$ in Table \ref{table:commuting_Killing}.
For Types I to VI, there are more than two commuting Killing vectors
in ${\cal C}(\xi)$, hence, the geodesic equations in
$(\orbit, \tilde{h})$ are integrable.  
As shown in the next section, the equations of motions for these strings
are not only integrable but also solved exactly. 
For the strings of Type VII, there is only one Killing vector
in ${\cal C}(\xi)$.
Nevertheless, the geodesic equations are solved exactly. This is due
to the existence of a Killing tensor in $(\orbit, \tilde{h})$.
We also solve the geodesic equation exactly.

\begin{table}[ht]
\caption{Inherited symmetry of $(\orbit,\tilde{h})$.
 $P_\mu~(\mu = t, x, y, z)$ is the generator of translation for
 $\mu$-direction. $L_i~(i = x, y, z)$ are the generators of rotation
 around $i$-axis. $K_i~(i = x, y, z)$ are the generators of Lorentz
 boosts for $i$-directions. $n$ is the number of commuting basis in
 ${\cal C}(\xi)$.} 
\begin{indented}
\item[] \begin{tabular}{@{}cccc}
 \br
 Type & tangential Killing vector $\xi$
      & basis of ${\cal C}(\xi)$ & $n$ \\
 \mr
 I   & $P_t + a L_z$  ($a\ne0$)
     & $P_t, P_z, L_z$ & 2 \\
 I   & $P_t$
     & $P_t, P_x, P_y, P_z, L_x, L_y, L_z$ & 3 \\
 I   & $L_z$
     & $P_t, P_z, L_z, K_z$ & 3 \\
 II  & $(P_t + P_z) + a L_z$  ($a\ne0$)
     & $P_t, P_z, L_z$ & 2 \\
 II  & $P_t + P_z$
     & $P_t, P_x, P_y, P_z, K_y + L_x, K_x - L_y, L_z$ & 3 \\
 III & $P_z + a L_z$  ($a\ne0$)
     & $P_t, P_z, L_z$ & 2 \\
 III & $P_z$
     & $P_t, P_x, P_y, P_z, L_z, K_x, K_y$ & 3 \\
 IV  & $P_z + a (K_y + L_z)$
     & $P_t - P_x, P_z, P_y + a (K_z - L_y), K_y + L_z$ & 2 \\
 V   & $P_z + a K_y$ ($a\ne0$)
     & $P_x, P_z, K_y$ & 2 \\
 V   & $K_y$ 
     & $P_x, P_z, L_y, K_y$ & 2 \\
 VI  & $K_y + L_z + a P_x$ ($a\ne0$)
     & $K_y + L_z + a P_x, P_t - P_x, P_z$ & 2 \\
 VII & $K_z + a L_z$ ($a\ne0$) 
     & $L_z, K_z$ & 1 \\
 \br
\end{tabular}
\end{indented}

\label{table:commuting_Killing}
\end{table}

\section{Solutions of cohomogeneity-one strings in Minkowski
spacetime} \label{sec:solutions}

\subsection{Type I}

The tangent Killing vector of this class is given as
\begin{equation}
 \xi = P_t + a L_{z},~~(a: \mbox{const.})
\end{equation}
where $P_t$ is the Killing vector of time translation and 
$L_z$ is that of rotation around $z$-axis.
In the conventional cylindrical coordinate
$(\bar{t}, \bar{\rho},\bar{\phi}, \bar{z})$ of Minkowski
spacetime, $\xi$ is written as
\begin{equation}
 \xi = \partial_{\bar{t}} + a \partial_{\bar{\phi}}, 
\end{equation}
and the norm of $\xi$ is
\begin{equation}
 f = |\xi|^2 = -(1 - a^2 \bar{\rho}^2).
\end{equation}
Then, $\xi$ is timelike in $\bar{\rho} < 1 / |a|$ and spacelike in 
$\bar{\rho} > 1 / |a|$. The surface $\bar{\rho} = 1 / |a|$ is called
light cylinder. Cohomogeneity-one strings of Type I inside the light
cylinder are the stationary rotating strings.
The constant $a$ represents the angular velocity of the rotation.

Here, we introduce a coordinate
$(t, \rho, \phi, z)
  = (\bar{t}, \bar{\rho}, \bar{\phi} - a \bar{t},\bar{z})$
so that $\xi = \partial_{t}$, i.e., one of the coordinates, say
$t$, is a coordinate along the orbits of $H$ which is generated by
$\xi$. 
In the new coordinate, the spacetime metric is
\begin{equation}
 g = -(1 - a^2 \rho^2)dt^2 + 2a\rho^2dtd\phi
     + d\rho^2 + \rho^2 d\phi^2 + dz^2 \label{eq:metric}
\end{equation}
and the norm of $\xi$ is
\begin{equation}
 f = - (1 - a^2 \rho^2).
\end{equation}
Then, the norm-weighted metric on the orbit space is calculated as
\begin{equation}
 \tilde{h}
 = - fg + \xi\xi
 = (1 - a^2 \rho^2) (d\rho^2 + dz^2) + \rho^2 d\phi^2.
\end{equation}

We solve the geodesic equations in $(\orbit, \tilde{h})$ with the action
\begin{eqnarray}
 S &= \int (\frac{\lagrangian}{N} + N) d\sigma, \label{eq:modified_action} \\
 \lagrangian &= (1 - a^2 \rho^2) (\rho'^2 + z'^2) + \rho^2 \phi'^2,
\end{eqnarray}
where $\sigma$ is a parameter of the geodesic curve,
$N$ is a function of $\sigma$ and the prime denotes the derivative
with respect to $\sigma$.
The action \eref{eq:modified_action} is invariant under the
transformations:
\begin{eqnarray}
 \sigma &\mapsto \tilde{\sigma} = \tilde{\sigma}(\sigma), \\
 N      &\mapsto \tilde{N} = \frac{d\sigma}{d\tilde{\sigma}}N.
\end{eqnarray}
Therefore, the function $N$ determines the parametrization of the
geodesic curve.
We should note that even though we fix the functional form of $N$,
there remains residual freedom of the parametrization: 
\begin{equation}
 \sigma \mapsto \tilde{\sigma} = \pm \sigma + \sigma_0.
 \label{eq:repara_freedom}
\end{equation}

The variation with respect to $\phi$ leads a conserved quantity related to
the $\phi$-independence of $\tilde{h}$:
\begin{equation}
 \frac{\rho^2 \phi'}{N} = L ~~(\mbox{const}). \label{eq:L}
\end{equation}
We also have a conserved quantity related to the $z$-independence of
$\tilde{h}$:
\begin{equation}
 \frac{1-a^2\rho^2}{N}z' = P ~~(\mbox{const}). \label{eq:p}
\end{equation}
The other variations lead
\begin{eqnarray}
 (1 - a^2 \rho^2) (\rho'^2 + z'^2) + \rho^2 \phi'^2 = N^2, \\
 \left\{
  \frac{1 - a^2 \rho^2}{N} \rho'
 \right\}'
 - \frac{1}{N}
   \left\{
    - a^2 \rho(\rho'^2 + z'^2) + \rho \phi'^2
   \right\}
   = 0.
\end{eqnarray}
By fixing the parametrization freedom as
\begin{equation}
 N = 1 - a^2 \rho^2,
\end{equation}
we obtain
\begin{equation}
 \rho'^2 = 1 - P^2 + a^2 L^2 - a^2 \rho^2 - \frac{L^2}{\rho^2}.
\end{equation}
This equation is readily integrated as
\begin{eqnarray}
 a^2 \rho^2(\sigma)
          = \alpha + \beta \cos 2a(\sigma + \sigma_0),\\
 \alpha := \frac{1 - P^2 + a^2 L^2}{2} \geq 0, \\
 \beta  := \sqrt{\alpha^2 - a^2 L^2},
\end{eqnarray}
where $\sigma_0$ is an integration constant.
We can set $\sigma_0$ to zero by using the residual reparametrization
freedom \eref{eq:repara_freedom}.
Then, the solution is written as
\begin{equation}
 a^2 \rho^2(\sigma) = \alpha + \beta \cos 2a \sigma,
 \label{eq:sol_I_rho}
\end{equation}
Using the solution, we can solve \eref{eq:L} and \eref{eq:p} as
\begin{eqnarray}
 \phi(\sigma)
 &= - a^2 L \sigma
   + \tan^{-1}\left[
                 \frac{aL}{\alpha + \beta} \tan a\sigma
              \right]
   + \phi_0, \label{eq:sol_I_phi}\\
 z(\sigma) &= P \sigma + z_0, \label{eq:sol_I_z}
\end{eqnarray}
where $\phi_0$ and $z_{0}$ are  constants.

The string solution, i.e., embedding of the world surface
$(\tau, \sigma) \mapsto (t, \rho, \phi, z)$, is given by
\eref{eq:sol_I_rho}, \eref{eq:sol_I_phi}, \eref{eq:sol_I_z} and 
$t = \tau$.
The solution has four integration constants:
$P, L, \phi_0$ and $z_0$. $P$ and $L$ determine the
shape of the string. However, $\phi_0$ and $z_0$ have no physical
meaning, because we can identify the solution of $\phi_0 \neq 0$ and
$z_0 \neq 0$ with that of $\phi_0 = z_0 = 0$ by the isometries in $(\M, g)$
\begin{eqnarray}
 \phi &\mapsto \phi + \phi_0, \\
 z    &\mapsto z + z_0,
\end{eqnarray}
where we should remember that the spacetime metric \eref{eq:metric} 
does not depend on $\phi$ and $z$.

\subsection{Types II to VI}
For Types II to VI, we can reduce the equations of motion to the
geodesic equations in the orbit space and solve them in the same manner
as in the case of type I.
We summarize the results in the Table \ref{table:solutions}. 

\begin{table}
 \caption{\label{table:solutions}
 We show tangent Killing vectors $\xi$, coordinates for the reduction,
 orbit space metrics $\tilde{h}$ and solutions of the geodesic
 equations. $P$, $Q$ and $L$ are the conserved quantities related to the 
 Killing vectors of $\tilde{h}$, and $C$ is that related to the Killing
 tensor of $\tilde{h}$.
 $(\bar{t}, \bar{x}, \bar{y}, \bar{z})$ and
 $(\bar{t}, \bar{\rho}, \bar{\phi}, \bar{z})$ are the Cartesian  
 coordinate and cylindrical coordinate of Minkowski spacetime,
 respectively.
 }
 \begin{indented}
  \item[]
 \begin{tabular}{@{}cl}
 \br
 Type & \\
 \mr
 I  & $\xi = P_t + a L_z,~
       (\bar{t}, \bar{\rho}, \bar{\phi} ,\bar{z})
        = (t, \rho, \phi + a t, z),~
	\tilde{h} = (1- a^2 \rho^2)(d\rho^2 +dz^2)+ \rho^2 d\phi$,\\
 {} & $z(\sigma) = P \sigma, ~
       a^2\rho^2(\sigma) = \alpha + \beta \cos 2a \sigma, ~
       \phi(\sigma)
                = - a^2 L \sigma
                  + \tan^{-1}\left[
                                 \frac{2aL}{\alpha + \beta}
                                 \tan a\sigma
                             \right]$\\
 {}       & $\alpha := (1 - P^2 + a^2 L^2)/2,
             ~\beta := \sqrt{\alpha^2 - a^2 L^2}$\\
 \mr
 II & $\xi = P_t + P_z + a L_z,~
            (\bar{t}, \bar{\rho}, \bar{\phi}, \bar{z})
              = (u + \frac{v}{2}, \rho, \phi + a u, u - \frac{v}{2})$,\\
 {} & $\tilde{h} = dv^2 - 2 a \rho^2 dv d\phi - a^2 \rho^2 d\rho^2$,\\
 {} & $v(\sigma) = a P \sigma,~
            a^2 \rho^2(\sigma)
                      = aP(L + \sqrt{L^2 -1} \cos2a\sigma),$\\
 {} & $\phi(\sigma)
                      = - a L \sigma
                        + \tan^{-1}\left[
                                     (L - \sqrt{L^2 - 1})\tan a \sigma
                                   \right]$\\
 \mr
 III & $\xi = P_z + a L_{z},~
             (\bar{t}, \bar{\rho}, \bar{\phi} ,\bar{z})
              =(t, \rho, \phi +a z, z),~
             \tilde{h} = (1 + a^2 \rho^2)(dt^2 - d\rho^2)
                         - \rho^2 d\phi^2,$ \\
 {}  & $t(\sigma) = Q \sigma,~
	a^2 \rho^2(\sigma) = \alpha + \beta \cos 2a \sigma,~
        \phi(\sigma)
                = a^2 L \sigma
                  + \tan^{-1}\left[
                                \frac{aL}{\alpha + \beta}
                                \tan a \sigma
                             \right]$,\\
 {}       & $\alpha:= (- 1 + Q^2 - a^2 L^2) / 2,~
             \beta := \sqrt{\alpha^2 - a^2 L^2}$\\
 \mr
 IV & $\xi = P_z + a (K_y + L_z),~
            (\bar{t}, \bar{x}, \bar{y}, \bar{z})
             = (\frac{a^2}{2} \lambda^2 u + v,
                -\bar{t} + u, a \lambda u, \lambda + w)$\\
 {}      & $\tilde{h} = (1 + a^2 u^2)(2 du dv - du^2) - a^2 u^2 dw^2$\\
 {}      & $u(\sigma) = P \sigma,~
            v(\sigma) = \frac{1 + P^2 + Q^2}{2P} \sigma
                         + \frac{a^2 P}{6} \sigma^3
                         - \frac{Q^2}{2a^2 P^3 \sigma},~
            w(\sigma) = Q \left(
                             \sigma
                             - \frac{1}{a^2 P^2 \sigma}
                           \right)$\\
 \mr
 V & $\xi = P_z + a K_y~(K_y: \mbox{timelike})$\\
 {}     & $(\bar{t}, \bar{x}, \bar{y}, \bar{z})
           = (y \sinh at, x, y \cosh at, z + t),~
           \tilde{h} = a^2 y^2 dz^2 - (1 - a^2 y^2) (dx^2 + dy^2)$ \\
 {}     & $x(\sigma) = P \sigma,~
          a^2 y^2(\sigma)
                     = \alpha
                       \pm \frac{1}{2}
                            \left(
                               e^{2 a \sigma} + \beta^2 e^{-2 a \sigma}
                            \right)$\\
 {}     & $z(\sigma) = Q \sigma
                       + \frac{1}{2a}
                         \ln\left|
                              \frac{e^{2 a \sigma} \pm (\alpha + Q)}
                                   {e^{2 a \sigma} \pm (\alpha - Q)}
                            \right|,
           ~\alpha := (1 + P^2 + Q^2)/2,
           ~\beta := \sqrt{\alpha^2 - Q^2}$\\
 {}     &\dotfill \\
 {}	& $\xi = P_z + a K_y~(K_y: \mbox{spacelike})$\\
 {}     & $(\bar{t}, \bar{x}, \bar{y}, \bar{z})
           = (t \cosh ay, x, t \sinh ay, z + y),~
           \tilde{h} = - a^2 t^2 dz^2 + (1 + a^2 t^2)(dt^2 - dx^2)$ \\
 {}     & $x(\sigma) = P \sigma,~
           -a^2 t^2(\sigma)
                     = \alpha
                       - \frac{1}{2}
                         \left(
                            e^{2a \sigma}
                            + \beta^2 e^{-2 a \sigma}
                         \right)$\\
 {}     & $z(\sigma) = Q \sigma
                       + \frac{1}{2a}
                         \ln\left|
                              \frac{e^{2 a \sigma} \pm (\alpha + Q)}
                                   {e^{2 a \sigma} \pm (\alpha - Q)}
                            \right|,~
           \alpha := (1 + P^2 + Q^2)/2,~
           \beta := \sqrt{\alpha^2 - Q^2}$ \\
 \mr
 VI & $\xi = P_x + a (K_y + L_z),~
       (\bar{t},\bar{x},\bar{y},\bar{z})
        = (\frac{a^2}{6}\lambda^3 + a u \lambda + v,
           - \bar{t} + \lambda, \frac{a}{2} \lambda^2 + u, w),$\\
 {}      & $\tilde{h} = (2 a u - 1)(du^2 + dw^2) + dv^2$\\
 {}      & $2 a u(\sigma) - 1 
                        = \frac{Q^2}{1 - P^2} + a^2 (1-P^2)\sigma^2,~
            v(\sigma) = \frac{P Q^2}{1 - P^2} \sigma 
                        + \frac{a^2 P (1-P^2)}{3}\sigma^3$,\\
 {}      & $w(\sigma) = Q \sigma$\\
 \mr
 VII & $\xi = K_z + a L_z~(K_z: \mbox{timelike})$\\
 {}  & $(\bar{t}, \bar{\rho}, \bar{\phi}, \bar{z})
              =(z \sinh t, \rho, \phi + a t, z \cosh t),~
             \tilde{h} = (z^2 - a^2 \rho^2)(d\rho^2 + dz^2)
                         + z^2 \rho^2 d\phi^2$ \\
 {}  & $z^2(\sigma)
        = C \pm \frac{1}{2}
                     \left(
                       e^{2 \sigma} + \beta^2 e^{-2 \sigma}
                     \right),~
        a^2 \rho^2(\sigma)
        = C + \beta \cos 2 a (\sigma + \sigma_0)$\\
 {}  & $\phi(\sigma)
        = \tan^{-1} \left\{
                      \frac{a L}{C + \beta} \tan a (\sigma + \sigma_0)
                    \right\}
          + \frac{a}{2}
            \ln\left|
                  \frac{e^{2\sigma} \pm (C + a L)}
                       {e^{2\sigma} \pm (C - a L)}
               \right|,~
         \beta := \sqrt{C^2 - a^2 L^2}$ \\
 {}       &\dotfill \\
 {}       & $\xi = K_z + a L_z~(K_z: \mbox{spacelike})$\\
 {}       & $(\bar{t}, \bar{\rho}, \bar{\phi}, \bar{z})
              =(t \cosh z, \rho, \phi + a z, t \sinh z),~
             \tilde{h} = (t^2 + a^2 \rho^2)(dt^2 - d\rho^2)
                         - t^2 \rho^2 d\phi^2$ \\
 {}       & $t^2(\sigma)
                = - C + \frac{1}{2}
                            \left(
                              e^{2 \sigma} + \beta^2 e^{-2 \sigma}
                            \right),~
             a^2 \rho^2(\sigma)
                = C + \beta \cos 2a(\sigma + \sigma_0),$\\
 {}       & $\phi(\sigma)
                = \tan^{-1}\left\{
                              \frac{a L}{C +  \beta}
                              \tan a(\sigma + \sigma_0)
                           \right\}
                  + \frac{a}{2}
                    \ln\left|
                           \frac{e^{2\sigma} - (C + a L)}
                                {e^{2\sigma} - (C - a L)}
                       \right|$,
             ~$\beta := \sqrt{C^2 - a^2 L^2}$ \\
 \br
 \end{tabular}
 \end{indented}
\end{table}

\subsection{Type VII}
The tangent Killing vector of Type VII string is
\begin{equation}
 \xi = K_z + a L_z,
\end{equation}
where $K_z$ is a Killing vector of the Lorentz boost 
along $z$-axis.
As in the case of Type I, we introduce a coordinate suitable for
the reduction.
In order to find such a coordinate, 
we use a combination of the Rindler coordinate 
and a cylindrical rotating coordinate in the form
\begin{equation}
 (\bar{t}, \bar{\rho}, \bar{\phi}, \bar{z})
 =(z \sinh t, \rho, \phi + a t, z \cosh t) \label{eq:rindler} , 
\end{equation}
such that
$\xi$ is written as $\xi = \partial_t$.
Since this coordinate covers only the part of 
Minkowski spacetime where $K_z$ is timelike, 
we use another coordinate 
\begin{equation}
 (\bar{t}, \bar{\rho}, \bar{\phi}, \bar{z})
 =(t \cosh z, \rho, \phi + a z, t \sinh z). \label{eq:rindler-like}
\end{equation}
in the spacelike regions of $K_z$. 
In this coordinate, $\xi$ is written as $\xi = \partial_z$.

\subsubsection{Timelike regions of $K_z$}
We take the coordinate \eref{eq:rindler}
in the timelike regions of $K_z$.
With respect to the coordinate, the spacetime metric is written as
\begin{equation}
 g = - (z^2 - a^2 \rho^2) dt^2 + 2a\rho^2 d\phi dt
     + d\rho^2 + \rho^2 d\phi^2 + dz^2,
\end{equation}
and the norm of the Killing vector $\xi$ is
\begin{equation}
 f = - (z^2 - a^2 \rho^2).
\end{equation}
Then, the metric $\tilde{h}$ on $\orbit$ is given as
\begin{equation}
 \tilde{h} = (z^2 - a^2 \rho^2)(d\rho^2 + dz^2) + z^2 \rho^2 d\phi^2.
\end{equation}
This metric admits a manifest Killing vector $\partial_{\phi}$
and  an irreducible
Killing tensor
\begin{equation}
 K = a^2 \rho^2 (z^2 - a^2 \rho^2) dz^2
     + z^2 (z^2 - a^2 \rho^2) d\rho^2
     + z^2 \rho^2 (z^2 + a^2 \rho^2)d\phi^2.
\end{equation}

In order to solve the geodesic equations,
we start from the action \eref{eq:modified_action} with
\begin{equation}
 \mathcal{L} = (z^2 - a^2 \rho^2)(\rho'^2 + z'^2) + z^2 \rho^2 \phi'^2.
\end{equation}
The existence of the Killing vector and the Killing tensor ensures 
two conserved quantities, say $L$ and $C$ respectively;
\begin{eqnarray}
 \fl L  = z^2 \rho^2 \frac{\phi'}{N}, \label{eq:const_kv}\\
 \fl 2C = a^2 \rho^2 (z^2 - a^2 \rho^2) \left(\frac{z'}{N}\right)^2
     + z^2 (z^2 - a^2 \rho^2) \left(\frac{\rho'}{N}\right)^2 
     + z^2 \rho^2 (z^2 + a^2 \rho^2) \left(\frac{\phi'}{N}\right)^2.
 \label{eq:const_kt}
\end{eqnarray}
Here, we should note that the prime does not represent the
differentiation with an affine parameter.
The geodesic tangent with an affine parameter is written as
${x^a}' / N$.
We solve \eref{eq:const_kv}, \eref{eq:const_kt} and the constraint
equation 
\begin{equation}
 (z^2 - a^2 \rho^2)(\rho'^2 + z'^2) + z^2 \rho^2 \phi'^2 = N^2.
 \label{eq:variation_N}
\end{equation}
By fixing the parametrization freedom as
\begin{equation}
 N = z^2 - a^2 \rho^2,
\end{equation}
we can separate the variables;
\begin{eqnarray}
 \rho'^2 &= 2 C - \frac{L^2}{\rho^2} - a^2 \rho^2,\\
 z'^2    &= - 2C + z^2 + \frac{a^2L^2}{z^2},
\end{eqnarray}
and solve the equations exactly as
\begin{eqnarray}
 z^2(\sigma)
  = C \pm \frac{e^{2 \sigma} + \beta^2 e^{-2 \sigma}}{2},
    \label{eq:sol_VIIt_z}\\
 a^2 \rho^2(\sigma)
  = C + \beta \cos 2 a (\sigma + \sigma_0),
    \label{eq:sol_VIIt_rho}\\
 \phi(\sigma)
  = \tan^{-1} \left\{
                \frac{a L}{C + \beta} \tan a (\sigma + \sigma_0)
              \right\}
    + \frac{a}{2}
        \ln\left|
               \frac{e^{2\sigma} \pm (C + a L)}
                    {e^{2\sigma} \pm (C - a L)}
           \right|,
	\label{eq:sol_VIIt_phi}\\
 \beta := \sqrt{C^2 - a^2 L^2},
\end{eqnarray}
where $\sigma_0$ is an integration constant.

\subsubsection{Spacelike regions of $K_z$}
With respect to the coordinate \eref{eq:rindler-like},
the metric $\tilde{h}$ on the orbit space is written as
\begin{equation}
\tilde{h} = (t^2 + a^2 \rho^2)(dt^2 - d\rho^2)
            - t^2 \rho^2 d\phi^2.
\end{equation}
This metric also admits a Killing vector $\partial_{\phi}$ and a Killing
vector
\begin{equation}
 K = a^2 \rho^2 (t^2 + a^2 \rho^2) dt^2
     + t^2 \rho^2 (t^2 - a^2 \rho^2) d\phi^2
     + t^2 (t^2 + a^2 \rho^2) d\rho^2. 
\end{equation}
These Killing vector and Killing tensor ensure the existence of 
two conserved quantities, and then the geodesic equations are
integrable. 
With a calculation similar to that used in deriving
solutions \eref{eq:sol_VIIt_z}, \eref{eq:sol_VIIt_rho},
\eref{eq:sol_VIIt_phi},
we obtain the exact solutions
\begin{eqnarray}
 t^2(\sigma) = - C + \frac{1}{2}
              \left(
                e^{2 \sigma} + \beta^2 e^{-2 \sigma}
              \right), \\
 a^2 \rho^2(\sigma) = C + \beta \cos 2a(\sigma + \sigma_0),\\
 \phi(\sigma)
     = \tan^{-1} \left\{
                      \frac{a L}{C +  \beta}
                      \tan a(\sigma + \sigma_0)
                  \right\}
        + \frac{a}{2}
          \ln\left|
               \frac{e^{2\sigma} - (C + a L)}
                     {e^{2\sigma} - (C - a L)}
             \right|,\\
 \beta := \sqrt{C^2 - a^2 L^2},
\end{eqnarray}
where $\sigma_0$ is an integration constant,
$L$ is a conserved quantity related to the Killing vector
$\partial_{\phi}$ and $C$ is the one related to the Killing tensor $K$.

\section{Conclusion}
We have shown that the Nambu-Goto equations of motion for all of the
cohomogeneity-one strings in Minkowski spacetime $(\mathcal{M}, g)$ are
integrable. 
The cohomogeneity-one string is a string whose world surface is
tangent to a Killing vector field $\xi$.
The Killing vector $\xi$ generates an one-parameter isometry group, say
$H$, which acts on the world surface. Then, the world surface has
symmetry due to $H$.
By virtue of the symmetry on the world surface,
the equations of motion reduces to the geodesic equations on
the orbit space $\orbit:=\mathcal{M}/H$ with a norm-weighted metric 
$\tilde{h}_{\mu\nu} := - \xi^2 g_{\mu\nu} + \xi_{\mu} \xi_{\nu}$.
We have investigated the integrability of these geodesic equations.

The integrability of the geodesic equations is related to the existence
of Killing vectors.
In the case of $(\orbit, \tilde{h})$, 
we have shown that the projections of the Killing vectors in
$(\mathcal{M}, g)$ which commute with $\xi$ are also Killing vectors in
$(\orbit, \tilde{h})$, i.e., the Killing vectors are inherited from
$(\mathcal{M}, g)$.
We have focused on the number of these inherited Killing vectors.

For the cohomogeneity-one strings of Types I to VI, 
we have found that there are more than two commuting Killing 
vectors in $(\orbit, \tilde{h})$.
Existence of two or more commuting Killing vectors
guarantees the integrability of the geodesic equations in
$(\orbit, \tilde{h})$ because $\dim \orbit = 3$.
Then, the geodesic equations for Types I to VI are integrable.
We have also found that the geodesic equations are solved exactly.
The exact solutions are shown in Table \ref{table:solutions}.

For the remaining cohomogeneity-one strings, i.e., Type VII,
there is only one inherited Killing vector.
However, we have found a Killing tensor in $(\orbit, \tilde{h})$. 
Existence of the Killing vector and the Killing tensor
leads two conserved quantities of the geodesic, 
and then the geodesic equations are integrable.
We have also solved the geodesic equations exactly.

\ack
This work is supported in part by Keio Gijuku Academic Development  
Funds (T.K.) and the Grant-in-Aid for Scientific Research No.19540305 
(H.I.).


\end{document}